\documentclass[aps,prd,reprint,groupedaddress]{revtex4-2}
 
\usepackage{bm}
\usepackage{amssymb}
\usepackage{epsfig}
\usepackage{slashed}
\usepackage{color}
\usepackage{mathtools}
\usepackage{cases}
\usepackage{booktabs}
\usepackage{amsmath}
\usepackage{url}

\usepackage{graphicx}
\usepackage{subfig}
\usepackage{ragged2e}
\usepackage{xcolor} 

\usepackage[
colorlinks=true,
filecolor=black,
anchorcolor=blue,
linkcolor=blue,
citecolor=cyan, 
urlcolor=cyan,
linktocpage=true,
plainpages=false,
breaklinks=true,
pdfstartview=FitH
]{hyperref}
\usepackage{orcidlink}
 
\def\dd{\mathrm{d}}
\def\tcos{\text{cos}}
\def\tsin{\text{sin}}
\def\A{\mathcal{A}}
\def\hOmega{\hat{\Omega}}

\def\hp{\hat{p}}
\def\mC{\beta}
\def\d{\text{d}}

\begin{document}
\title{Detecting Chiral Gravitational Wave Background with a Dipole Pulsar Timing Array}


\author{Baoyu Xu$^{1,3}$}
\email[]{xuby@bao.ac.cn}

\author{Hanyu Jiang$^{1,3}$}

\author{Rong-Gen Cai$^{4}$}

\author{Misao Sasaki$^{5,6,7,8}$}

\author{Yun-Long Zhang$^{1,2}$}
\email[]{zhangyunlong@nao.cas.cn}

\affiliation{$^1$National Astronomical Observatories, Chinese Academy of Sciences, Beijing 100101, China}

\affiliation{$^2$School of Fundamental Physics and Mathematical Sciences, Hangzhou Institute for Advanced Study, University of Chinese Academy of Sciences, Hangzhou 310024, China}

\affiliation{$^{3}$School of Astronomy and Space Science, University of Chinese Academy of Sciences, Beijing 100049, China
}

\affiliation{$^{4}$
Institute of Fundamental Physics and Quantum Technology, Ningbo University, Ningbo 315211, China}


\affiliation{$^{5}$
Kavli Institute for the Physics and Mathematics of the Universe (WPI), The University of Tokyo, Chiba 277-8583, Japan}

\affiliation{$^{6}$
Asia Pacific Center for Theoretical Physics, Pohang 37673, Korea}

\affiliation{$^{7}$
Leung Center for Cosmology and Particle Astrophysics, National Taiwan University, Taipei 10617, Taiwan}

\affiliation{$^{8}$
Center for Gravitational Physics and Quantum Information, Yukawa Institute for Theoretical Physics, Kyoto University, Kyoto 606-8502, Japan}

\begin{abstract}
The pulsar timing array (PTA) is a powerful technique for detecting nanohertz gravitational wave backgrounds (GWBs). However, conventional PTAs lack sensitivity to parity violation in the GWB. In this work, we propose a dipole pulsar timing array system (dPTA). By deriving the overlap reduction functions (ORFs) from the cross-correlation of timing signals, we find that this system exhibits sensitivity to chiral GWBs in the nanohertz regime. Furthermore, through numerical calculations of its sensitivity curves, we demonstrate that the dPTA extends the detectable frequency range of PTAs for GWBs from the nanohertz to the microhertz regime.
\end{abstract}


\maketitle
\tableofcontents
\allowdisplaybreaks

\section{Introduction}

Recently, significant breakthroughs have been  achieved in detecting the  gravitational wave backgrounds (GWB) using pulsar timing array (PTA)~\cite{NANOGrav:2023gor,NANOGrav:2023icp,NANOGrav:2023hvm,NANOGrav:2023hde,Reardon:2023gzh,EPTA:2023sfo,EPTA:2023fyk,Miles:2024seg,Taylor:2013vha,Xu:2023wog,Bi:2023tib,Ellis:2023dgf}. 
The GWB carries valuable information about the early universe and serves as a key probe of its properties~\cite{Allen:1996vm,Caprini:2018mtu,Romano:2016dpx,Renzini:2022alw}. 
In particular, some cosmological and  astrophysical processes may induce GWB exhibiting circular polarization, which is referred to as chiral GWB. Cosmological chiral GWB can be produced by a wide variety of mechanisms, such as inflation~\cite{Alexander:2004us,Li:2020xjt,Li:2021wij,Cai:2021uup,Li:2024fxy,Cai:2016ihp,Takahashi:2009wc}, phase transitions~\cite{Cornwall:1997ms,Vachaspati:2001nb,Kahniashvili:2005qi,RoperPol:2019wvy,Ellis:2020uid} and axionic mechanisms~\cite{Machado:2018nqk,Ding:2024,Xu:2024kwy,Garriga:2025uko}. 
Moreover, astrophysical chiral GWB can be produced by shot noise fluctuations in the number of gravitational wave (GW) sources~\cite{ValbusaDallArmi:2023ydl} and precessing binary black hole mergers~\cite{Leong:2025raf}.
Therefore, probing the chiral GWB would have significant implications for fundamental physics.

Probing the chiral GWB requires not only measuring its amplitude but also determining the asymmetry between the right- and left-handed polarization modes.
For ground-based detectors, the current sensitivity is insufficient to constrain the chirality of chiral GWB, and further measurements are still required~\cite{Martinovic:2021hzy,Jiang:2022uxp,Omiya:2023rhj}.
Space-based GW detection projects such as LISA~\cite{Barausse:2020rsu,LISA:2022kgy}, Taiji~\cite{Ruan:2018tsw}, and TianQin~\cite{TianQin:2015yph,Luo:2025ewp} make it possible to identify chiral GWB in millihertz (mHz)  frequency band by  constructing network~\cite{Seto:2006dz,Seto:2020zxw,Ruan:2020smc,Wang:2021uih,Liu:2022umx,Chen:2024xzw,Chen:2024ikn,Chen:2024fto,Su:2025nkl}. However, in nanohertz (nHz) frequency band, PTA is not sensitive to the chirality of the GWB signals in isotropic background~\cite{Kato:2015bye,Belgacem:2020nda,Sato-Polito:2021efu}, although astrometry has the potential to detect chiral GWB in the nHz frequency band~\cite{Liang:2023pbj,Jaraba:2025hay,Caliskan:2023cqm}. 
On the other hand, current PTA projects cannot cover the microhertz ($\mu$Hz) frequency band, therefore, a method capable of detecting the GWB in the nHz$\sim\mu$Hz range is needed.

In this work, we propose the dipole system based on PTA, which can be named as dipole pulsar timing array (dPTA). It not only exhibits sensitivity to the chirality of the GWB, but also extends the detectable frequency range from nHz to $\mu$Hz.
Specifically, we first derive the nonzero overlap reduction functions (ORFs) associated with the GWB intensity and chirality, indicating that the dPTA can respond to the chiral GWB in an isotropic background. Then, through signal-to-noise ratio (SNR) analysis, we analytically derive and numerically compute the sensitivity curves for the dPTA. 

The structure of this paper is as follows. In section~\ref{section signal}, we construct the dPTA based PTA and present the form of GW signal. We briefly review the theoretical framework of the chiral GWB and PTA in section~\ref{section GWB}. Then, we use cross correlation statistics to show dPTA's responses to chiral GWB, which reveals nonzero ORFs for both intensity and chirality in section~\ref{section ORF}. In section~\ref{section sensitivity curve}, we compute the sensitivity curves of dPTA  using the power-law integrated curve method under the condition of maximized SNR. To eliminate parameter mixing, we then evaluate the effective SNRs for different Stokes parameters and obtain the sensitivity curves for them. We show that the dPTA can extend the PTA detection band from nHz to $\mu$Hz.  The conclusions and discussions are presented in the last section~\ref{section conclusion}. Throughout this paper, we adopt the metric signature as $(-,+,+,+)$, and set $c=\hbar=1$.

\section{dPTA System and Chiral GWB}\label{section signal}

GWB can be described as the superposition of plane waves from all possible directions and frequencies~\cite{Caprini:2018mtu,Husa:2009zz}, 
\begin{equation}\label{GW fourier}
    h_{ij} = \sum_{\lambda=+,\times} \int_{-\infty}^\infty df \int_{S^2} d\hat{\Omega}\; \tilde{h}_{\lambda}(f, \hat{\Omega})\, e^{\lambda}_{ij}(\hat{\Omega})e^{ i 2\pi f (t - \hat{\Omega} \cdot \vec{x})},
\end{equation}
where $e^{\lambda}_{ij}(\hOmega)$ is the polarization tensor with  $\lambda=+,\times$,  $\tilde{h}_{\lambda}(f, \hat{\Omega})$ represents the Fourier transform of the plane waves. For a homogeneous and isotropic GWB, its ensemble average is given by~\cite{Seto:2020zxw}:
\begin{align}\label{average}
 &\langle \tilde{h}_{\lambda}(f,\hat\Omega)\tilde{h}_{\lambda'}^{*}(f',\hat\Omega')\rangle\nonumber\\
 =   &\frac{1}{8\pi}\delta(f-f')\delta(\hOmega-\hOmega') 
     \left[\begin{array}{cc}
     I(f) & -iV(f) \\
     iV(f) & I(f)
     \end{array}\right].
\end{align}
Here, $I$ and $V$ represent the Stokes parameters. Using the circular polarization decomposition $e_{ij}^{\A}=(e^{+}_{ij}\pm ie^{\times}_{ij})/\sqrt{2}$ and $h_{\A}=(h_{+}\mp ih_{\times})/\sqrt{2}$ (where $\A=R,L$) to describe the GWB~\cite{Seto:2006dz,Seto:2020zxw}, the correlation in above equation becomes $\langle \tilde{h}_{\A}\tilde{h}_{\A}^{*} \rangle=\frac{S_{\A}(f)}{8\pi}\delta(f-f')\delta(\hOmega-\hOmega')\delta_{\A\A'}$.  
Here, $S_{\A}(f)$ represents the spectral density of right- or left-handed modes. Then, the Stokes parameters can be expressed as $I(f)=\frac{1}{2}\left[S_{R}(f)+S_{L}(f)\right]$ and $V(f) =\frac{1}{2}\left[S_{R}(f)-S_{L}(f)\right].$ 
This implies that $I$ represents the intensity of the GWB, while $V$ represents its chirality. 


\subsection{Chiral GWB and PTA}
\label{section GWB}

Detecting the stochastic background most effectively relies on utilizing the statistical properties of the signal. 
The averaged cross-correlation is thus employed to extract the signal induced by the GWB.  For signal $z_a(t)$, the averaged correlation is calculated as
\begin{equation}\label{correlationS}
\begin{split}
& \langle z_{ab}(\tau)\rangle = \int \dd t \langle z_{a}(t)z_{b}(t+\tau)\rangle \\
    &= \iint \dd f \dd f' \langle \tilde{z}_{a}(f) \tilde{z}^{*}_{b}(f') \rangle \delta_T(f-f')e^{-i 2\pi f' \tau} .
\end{split}
\end{equation}
Here, $a$ and $b$ denote different signals. $\tilde{z}_a(f)$ is the Fourier transform of  $z_a(t)$  while $\delta_{T}(f-f')$ is the finite-time approximation to the Dirac delta function with $\delta_{T}(0)=T$. Specifically, by using the ensemble average in Eq.~\eqref{average}, their correlation can be expressed as $\langle \tilde{z}_{a}(f) \tilde{z}_{b}^{*}(f') \rangle
= \frac{1}{2}\delta(f-f')\left[I(f)\Gamma_{ab}^{I}(f)+V(f)\Gamma_{ab}^{V}(f)\right]$. Then, the above Eq.~\eqref{correlationS} can be rewritten as
\begin{equation}\label{cor}
\begin{split}
\langle z_{ab}(\tau)\rangle= \frac{T}{2} \int\dd f&\big[I(f)\Gamma_{ab}^{I}(f)\tcos(2\pi f \tau)
\\
&-V(f) i \Gamma_{ab}^{V}(f)\tsin(2\pi f \tau)\big],
\end{split}
\end{equation}
where $\Gamma_{ab}^{I}(f)$ and $\Gamma_{ab}^{V}(f)$ are the ORFs for Stokes parameters $I$ and $V$, respectively.

For PTA, its signal from pulsar-a can be expressed as~\cite{Anholm:2008wy}
\begin{align}\label{redshift}
    z_{a}(t) = \sum_{\lambda}\int \dd f \int_{S^{2}} \dd \hOmega \tilde{h}_{\lambda}(f, \hat{\Omega}) \, \mathcal{E}_{a}({\hOmega})F_{a}^{\lambda}(\hOmega) e^{i2\pi f t}.
\end{align}
Here, we have $F^{\lambda}_{a}(\hOmega)=\hat{p}_{a}^{i}\hat{p}_{a}^{j}e^{\lambda}_{ij}/2(1+\hOmega\cdot\hat{p}_{a})$ and $\mathcal{E}_{a}({\hOmega})=(e^{i2\pi f L_{a}(1+\hOmega\cdot\hat{p}_{a})}-1)$ with the distance from earth to the pulsar $L_{a}$. Thus, the ORFs can be calculated as
\begin{align}
    \Gamma^{I}_{ab} &= \int_{S^{2}}  \frac{\dd\hOmega}{4\pi}\mathcal{E}_{a}(\hOmega)\mathcal{E}^{*}_{b}(\hOmega)\left(F^{+}_{a}F^{*+}_{b}+F^{\times}_{a}F^{*\times}_{b}\right),
    \\
    i\Gamma_{ab}^{V} &= \int_{S^{2}}  \frac{\dd\hOmega}{4\pi}\mathcal{E}_{a}(\hOmega)\mathcal{E}^{*}_{b}(\hOmega)\left(F^{+}_{a}F^{*\times}_{b}-F^{\times}_{a}F^{*+}_{b}\right).\label{ptaVORFs}
\end{align}

Although the PTA signal can be expressed in the form of Eq.~\eqref{cor},  it still shows no response to the Stokes parameter $V$. That is because $i\Gamma^{V}_{ab}=0$. Specifically, when the distance from each pulsar to Earth is much longer than the GW wavelength  ($fL_{a},fL_{b}\gg1$), we can apply the approximation 
$\mathcal{E}_{a}(\hOmega)\mathcal{E}^{*}_{b}(\hOmega)=(e^{i2\pi f L_{a}(1+\hOmega\cdot\hat{p}_{a})}-1)(e^{-i2\pi f L_{b}(1+\hOmega\cdot\hat{p}_{b})}-1)\simeq 1$
and the $i\Gamma^{V}_{ab}$ can be rewritten as $i\Gamma_{ab}^{V}(f)=\int_{S^{2}}\frac{\dd\hOmega}{4\pi}\left(F^{+}_{a}F^{*\times}_{b}-F^{\times}_{a}F^{*+}_{b}\right)$. Since the ORFs of PTA depend only on the angle between the two pulsars, it is convenient to adopt a coordinates system where $\hat{p}_{a}=(0,0,1)$, $\hat{p}_{b}=(\tsin\alpha,0,\tcos\alpha)$ and $\hOmega = (\text{sin}\theta\text{cos}\phi,\text{sin}\theta\text{sin}\phi,\text{cos}\theta)$, which gives us
\begin{equation}
    \begin{split}\label{gammaV}
       i \Gamma_{ab}^{V}(f) &= \int^{\pi}_{0}\dd \theta\frac{\tsin\theta}{8\pi}(1-\tcos\theta) \int^{\pi}_{-\pi} \dd \phi ~\tsin\phi~ \Theta(\phi,\theta),
    \end{split}
\end{equation}
where
$\Theta(\phi, \theta)\equiv\frac{\tsin\alpha (\tcos\alpha~\tsin\theta~-\tsin\alpha~\tcos\theta~\tcos\phi)}{1+\tsin\alpha~\tsin\theta~\tcos\phi+\tcos\alpha~\tcos\theta}$, and it is an even function with respect to $\phi$. Consequently, the integral over $\phi$ vanishes, resulting to  $i \Gamma_{ab}^{V}(f)=0$. This explains why PTA exhibits no response to  $V$.
Although PTA is insensitive to $V$ in an isotropic background, as demonstrated in ~\cite{Kato:2015bye}, we will show that the dPTA can effectively compensate for this limitation.

\color{black}

\subsection{Dipole PTA System}\label{section ORF}

\begin{figure}[h]
\captionsetup{justification=raggedright,singlelinecheck=false}
\includegraphics[scale=0.3]{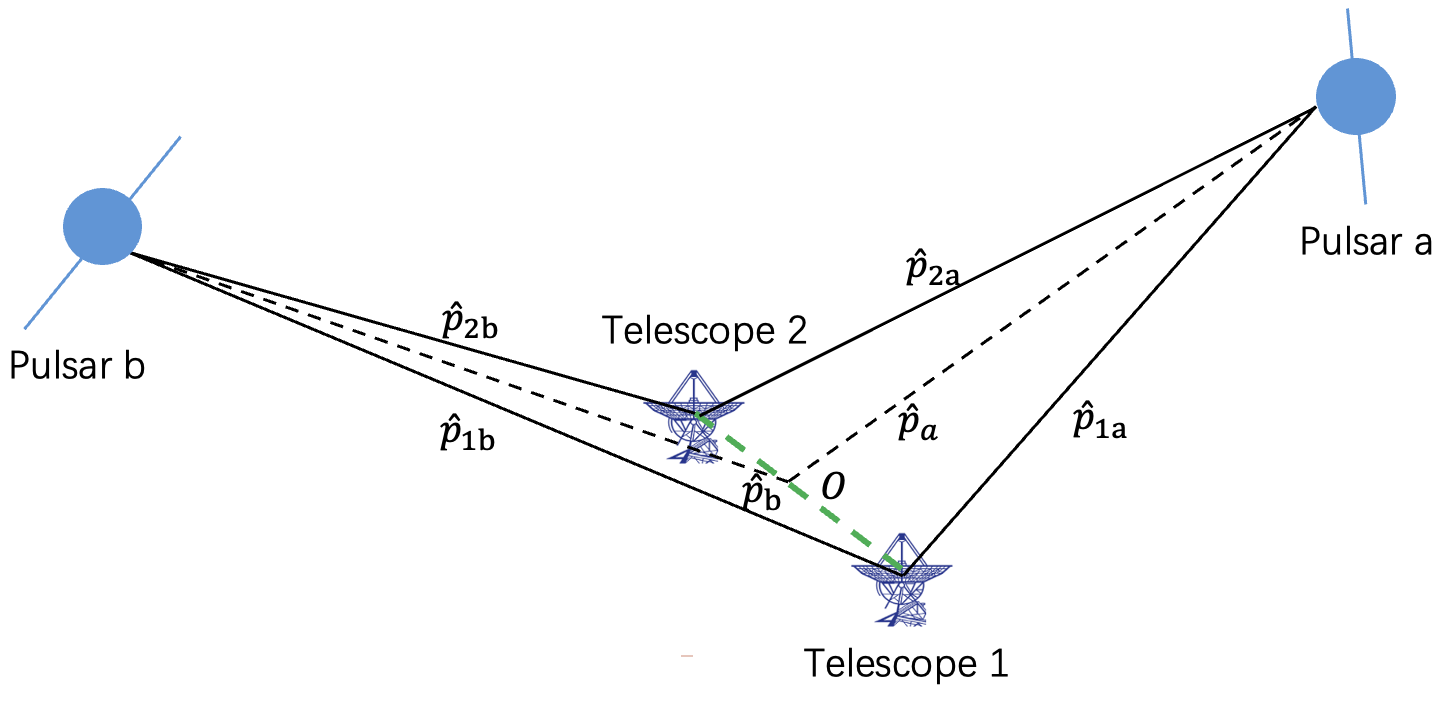}
\caption{The diagram for  dipole pulsar timing (dPTA).}
\label{DRTT}
\end{figure}

The alternative system we propose is the dipole PTA, illustrated in Fig.~\ref{DRTT}. Two radio telescopes, separated by a distance $D$, probe the same pulsar. 
Thus, we can obtain the chiral GWB signal by extracting the difference between the timing residuals of the two telescopes.

Specifically, for a single detector, the redshift induced by GWB is the same as that of PTA, which can be expressed as Eq.~\eqref{redshift}. 
We define the dPTA signal from pulsar-a $s_{a}(t)$ as the redshift difference between the two detectors induced by the GWB: $s_{a}(t)= z_{1a}(t)-z_{2a}(t)$. It is convenient to use the emission time $t$. When the distance from Pulsar-a to the system satisfies $L_{1a},L_{2a} \gg D$, we can neglect higher order terms,  so that the dPTA signal is written as $s_{a}(t)=\int_{S^2}\dd\hat{\Omega} \Delta H_{ij}^{a}\mathcal{F}_{a}$.
Here, $\mathcal{F}_{a}=\hat{p}_{a}^{i}\hat{p}_{a}^{j}/2(1+\hOmega\cdot\hat{p}_{a})$, where the unit vector $\hat{p}_{a}$ pointing from the $O$ to Pulsar-a, and 
\begin{align}\label{new perturbation}
\Delta H_{ij}^{a} = &\sum_{\lambda=+,\times} \int_{-\infty}^\infty \dd f \tilde{h}_{\lambda}(f, \hat{\Omega}) \, e^{\lambda}_{ij}(\hat{\Omega})e^{i2\pi f (t-L_{a}\hOmega\cdot\hat{p}_{a})} \nonumber\\
 &\times\left[e^{i 2\pi f L_{2a}(1 + \hat{\Omega} \cdot \hat{p}_{2a})} -e^{i 2\pi f L_{1a}(1 + \hat{\Omega} \cdot \hat{p}_{1a})} \right].
\end{align}

Then, by using the averaged correlation in Eq.~\eqref{correlationS}, 
the ORFs for dPTA can be culatated as
\begin{align}
  & \widetilde{\Gamma}_{ab}^{I}(f) = \int \frac{\dd \hOmega}{4\pi}  \frac{\hat{p}_{a}^{i} \hat{p}_{a}^{j} \hat{p}_{b}^{l} \hat{p}_{b}^{m}(e^{+}_{ij}e^{+}_{lm}+e^{\times}_{ij}e^{\times}_{lm})}{(1 + \hat{\Omega} \cdot \hat{p}_{a})(1 + \hat{\Omega} \cdot \hat{p}_{b})}\mathcal{E}_{a b}(f, \hOmega), \label{GammaI}
    \\
  & i\widetilde{\Gamma}_{ab}^{V}(f) = \int \frac{\dd \hOmega}{4\pi}  \frac{\hat{p}_{a}^{i} \hat{p}_{a}^{j} \hat{p}_{b}^{l} \hat{p}_{b}^{m}(e^{+}_{ij}e^{\times}_{lm}-e^{\times}_{ij}e^{+}_{lm})}{(1 + \hat{\Omega} \cdot \hat{p}_{a})(1 + \hat{\Omega} \cdot \hat{p}_{b})} \mathcal{E}_{a b}(f, \hOmega).
  \label{GammaV}
\end{align}
Here, we have introduced the notation
$\mathcal{E}_{ab}(f, \hOmega) \equiv  
\tsin[\pi f D(\hat{p}_{a}+\hOmega)\cdot\hat{D}]\,
\tsin[\pi f D(\hat{p}_{b}+\hOmega)\cdot\hat{D}],$
where $\hat{D}$ denotes the unite direction vector of the baseline.
Unlike in PTA, the ORFs for dPTA depend on both space direction and frequency.

Notice that in Eq.\eqref{average}, the Stokes parameter $V$ induces the imaginary off-diagonal term. Therefore, the ORF $\widetilde{\Gamma}_{ab}^{V}(f)$ in Eq.~\eqref{GammaV} turns out to be imaginary and non-zero if $a\neq b$. This characteristic of $\widetilde{\Gamma}_{ab}^{V}(f)$ is consistent with the usual decomposition for chiral GWB ~\cite{Seto:2006hf,Seto:2020zxw,Chen:2024fto,Chen:2024ikn}. 
 Although the $\langle \tilde{s}_{a}(f) \tilde{s}_{b}^{*}(f') \rangle$ has an imaginary component, in the time domain the ensemble average  $\langle s_{ab}(\tau) \rangle$ is still real-valued with the contribution of the $V$. This occurs because the odd function $V$ is retained by the odd component in $e^{-i2\pi f \tau}$,  resulting in: 
$\langle s_{ab}(\tau)\rangle= \frac{T}{2} \int\dd f [I(f)\widetilde{\Gamma}_{ab}^{I}(f)\tcos(2\pi f \tau) -V(f) i \widetilde{\Gamma}_{ab}^{V}(f)\tsin(2\pi f \tau) ]$.

\begin{figure}[h]
\captionsetup{justification=raggedright,singlelinecheck=false}
\includegraphics[scale=0.20]{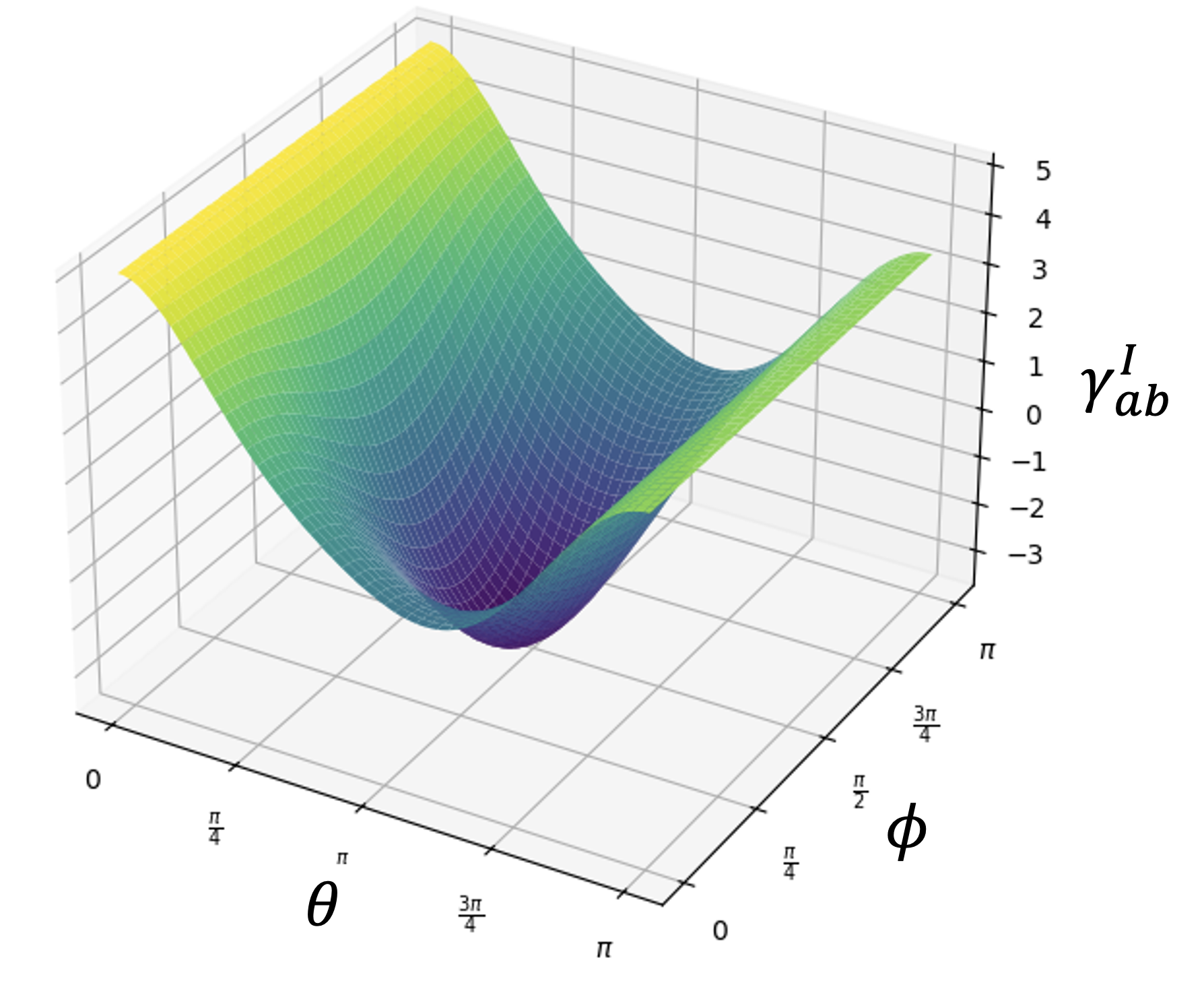}
\caption{\justifying
The normalized ORF $\gamma^{I}_{ab}$ is plotted based on Eq.~\eqref{GammaI} and Eq.~\eqref{GammaV1}. It represents the spatial dependence of response for $I$. We set the baseline $\hat{D}$ along the x-axis, fix $\hp_{a}$ along the z-axis, and let the direction of $\hp_{b}$ be arbitrary. Since the system is symmetric with respect to the $x\text{-}z$ plane, we only plot $\phi\in[0,\pi]$.}
\label{3DI}
\end{figure}
\begin{figure}[h]
\captionsetup{justification=raggedright,singlelinecheck=false}
\includegraphics[scale=0.20]{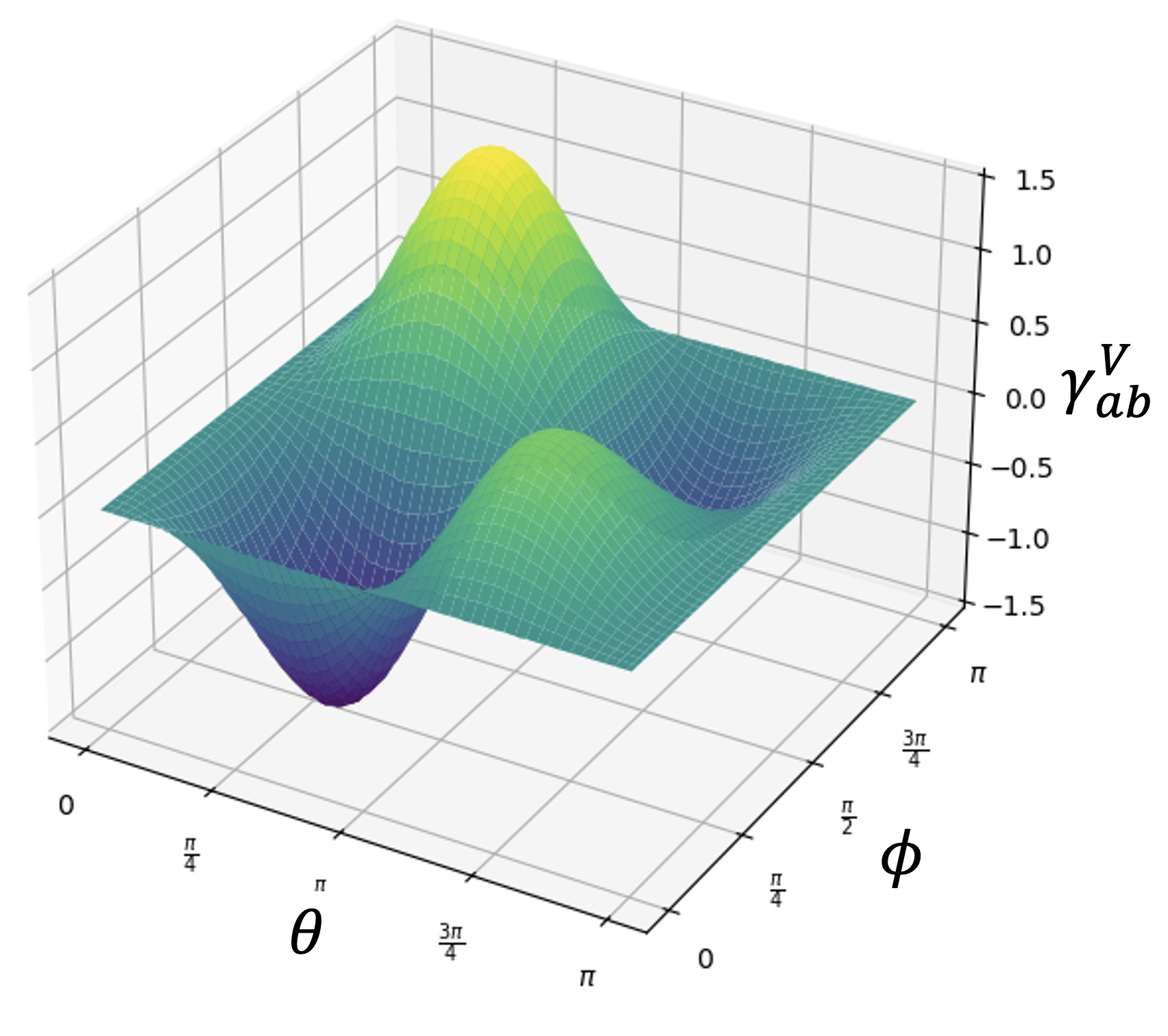}
\caption{\justifying
The normalized ORF $\gamma^{V}_{ab}$ is plotted based on Eq.~\eqref{GammaV} and Eq.~\eqref{GammaV1}. It represents the spatial dependence of response for $V$. The configuration of $\hp_a$, $\hp_b$ and $\hat{D}$ is the same as that in Fig.~\ref{3DI}. When $\hp_a$, $\hp_b$, and $\hat{D}$ are not coplanar, $\gamma_{ab}^{V}$ does not vanish upon integration over the whole sky, indicating that the dPTA is sensitive to chirality of the GWB.}
\label{3DV}
\end{figure}

When the GW wavelength is much larger than the baseline $D$, the $\pi fD$ can be regarded as an infinitesimal quantity,  implying that $\tsin\big[\pi f D (\hat{p}_{a}+\hOmega)\cdot\hat{D} \big]
\simeq\pi f D (\hat{p}_{a}+\hOmega)\cdot\hat{D}$. Therefore, we can introduce the normalization factor $4\pi/\tsin^{2}(\pi f D)$ to extract the spatially dependent part of the ORFs. The normalized ORFs can be written as
\begin{align}
   \gamma_{ab}^{I}&=\frac{4\pi}{\tsin^{2}(\pi f D)}\widetilde{\Gamma}_{ab}^{I}(f),\quad 
   \gamma_{ab}^{V}=\frac{4\pi}{\tsin^{2}(\pi f D)}i\widetilde{\Gamma}_{ab}^{V}(f). \label{GammaV1}
\end{align} 
Since the baseline defines a specific direction $\hat{D}$, the ORFs are determined by the relative positions of the pulsars, characterized by their angular coordinates $\theta$ and $\phi$. We set the baseline along the $x$-axis and $\hp_{a}$ along the $z$-axis, studying the spatial dependence of $\hp_{b}$ relative to the $\hp_{a}$. The results of the normalized ORFs are plotted in Fig.~\ref{3DI} and Fig.~\ref{3DV}.

When $\hat{p}_{a}$, $\hat{p}_{b}$ and $\hat{D}$ are not coplanar, the detection system breaks parity and thus exhibits a natural response to $V$. For both $I$ and $V$, they have nonzero angle dependent ORFs, indicating that the system responds to both components.

\section{Sensitivity Curves of Chiral GWB}
\label{section sensitivity curve}

Since most GWB spectra exhibit a power-law behavior, the corresponding sensitivity curves can be obtained using the power-law integrated curve method. This frequentist detection method relies on SNR analysis. 

In our framework, the independent timing of each pulsar represents an individual detector, while the cross-correlation between different pulsars can be viewed as a measurement within a detector network. Thus, the maximum SNR is given by~\cite{Thrane:2013oya}
\begin{equation}\label{SNR}
    \text{SNR}^{2} = 2T \int^{f_{\text{max}}}_{f_{\text{min}}} \dd f \sum_{a,b}\frac{\left[\widetilde{\Gamma}_{ab}^{I}(f)I(f)+\widetilde{\Gamma}_{ab}^{ V}(f)V(f)\right]^{2}}{P(f)P(f)}. 
\end{equation}
Here, $\sum_{a,b}$ denotes the summation over all possible pulsar pairs and $P(f)$ is the nosie. We assume that the pulsars have identical white noise and white noise power spectrum of the redshift signal can be written as~\cite{hobbs2010international,EPTA:2023akd}
\begin{equation}
    P(f) = (2\pi f)^{2}\cdot(2 \Delta t \sigma^{2}).
\end{equation}
Here, $\sigma$ denotes the root mean square measurement variance, and $\Delta t$ represents the sampling interval. In actual observations, what we obtain is the time of arrival (TOA). For TOA, $2 \Delta t \sigma^{2}$ corresponds to the time noise power spectrum. Since our previous analysis was based on the redshift, we need to convert it from the time noise power spectrum into the redshift noise power spectrum. This transformation introduces additional factor of $(2\pi f)^{2}$.

\begin{figure}[h]
\captionsetup{justification=raggedright,singlelinecheck=false}
\includegraphics[scale=0.45]{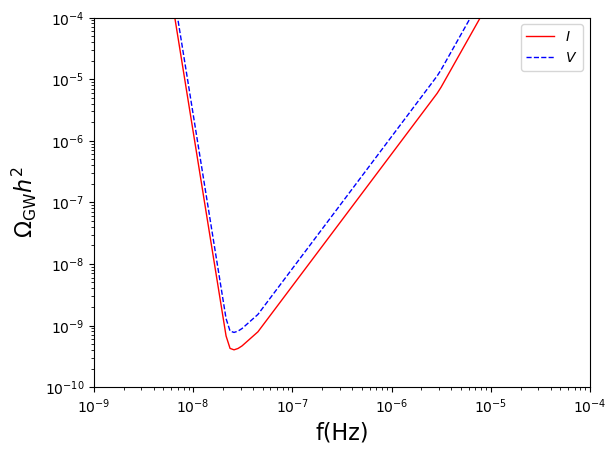}
\caption{
\justifying
Power-law sensitivity curves for the Stokes parameters $I$ (red line) and $V$ (blue line). We select baseline $D=1 \text{AU}$ and observation time $T=1.5$yr. The dPTA shows comparable sensitivity to both Stokes parameters. }
\label{IV}
\end{figure}

Since the total SNR contains the responses to both $I$ and $V$, joint analysis using the Fisher information matrix is required  to properly evaluate the SNR for different parameters and to avoid parameter coupling. By inverting the Fisher matrix, we obtain the covariance matrix, from which the effective SNR  of each parameter can be determined. 
Assuming the uniform distribution of pulsars, the effective SNRs can be given by
$\text{SNR}_{I}^{2} = 
    2T\int^{f_{\text{max}}}_{f_{\text{min}}} \dd f \frac{I(f)^{2}}{P(f)^{2}} \widetilde{\Gamma}^{ I}_{\text{eff}}(f)^2$,
$\text{SNR}_{V}^{2} = 
    2T\int^{f_{\text{max}}}_{f_{\text{min}}} \dd f  \frac{V(f)^{2}}{P(f)^{2}}  \widetilde{\Gamma}^{ V}_{\text{eff}}(f) ^2$,    
where $\widetilde{\Gamma}^{I}_{\text{eff}}(f)^2=\frac{N(N-1)}{2}\big(\overline{\widetilde{\Gamma}_{ab}^{ I}(f)^2}-\overline{\widetilde{\Gamma}_{ab}^{I}(f) \widetilde{\Gamma}_{ab}^{V}(f)}^{2}/\overline{\widetilde{\Gamma}_{ab}^{V}(f)^2}\big)$, and $\widetilde{\Gamma}^{V}_{\text{eff}}(f)^2=\frac{N(N-1)}{2}\big(\overline{\widetilde{\Gamma}_{ab}^{ V}(f)^2}-\overline{\widetilde{\Gamma}_{ab}^{I}(f) \widetilde{\Gamma}_{ab}^{ V}(f)}^{2}/\overline{\widetilde{\Gamma}_{ab}^{I}(f)^2}\big)$.
Here, we choose number of pulsars $N=100$, which denotes the number of available millisecond pulsars~\cite{IPTAweb,Perera:2019sca}. 
Based on the SNR formula~\eqref{SNR}, we can use the power-law integrated curves method described in~\cite{Thrane:2013oya} to obtain the sensitivity curves.
We set  the effective parameter  $\text{SNR}_{I,V} =1 $, measurement variance $\sigma=0.1\text{ns}$ and sampling interval $\Delta t=10^{3}\text{s}$. For dPTA system with a baseline $D=1 \text{AU}$, the sensitivity curves for Stokes parameters $I$ and $V$ are shown in Fig.~\ref{IV}. It can be seen that dPTA exhibits comparable sensitivity to both $I$ and $V$. 

\begin{figure}[h]
\captionsetup{justification=raggedright,singlelinecheck=false}
\includegraphics[scale=0.5]{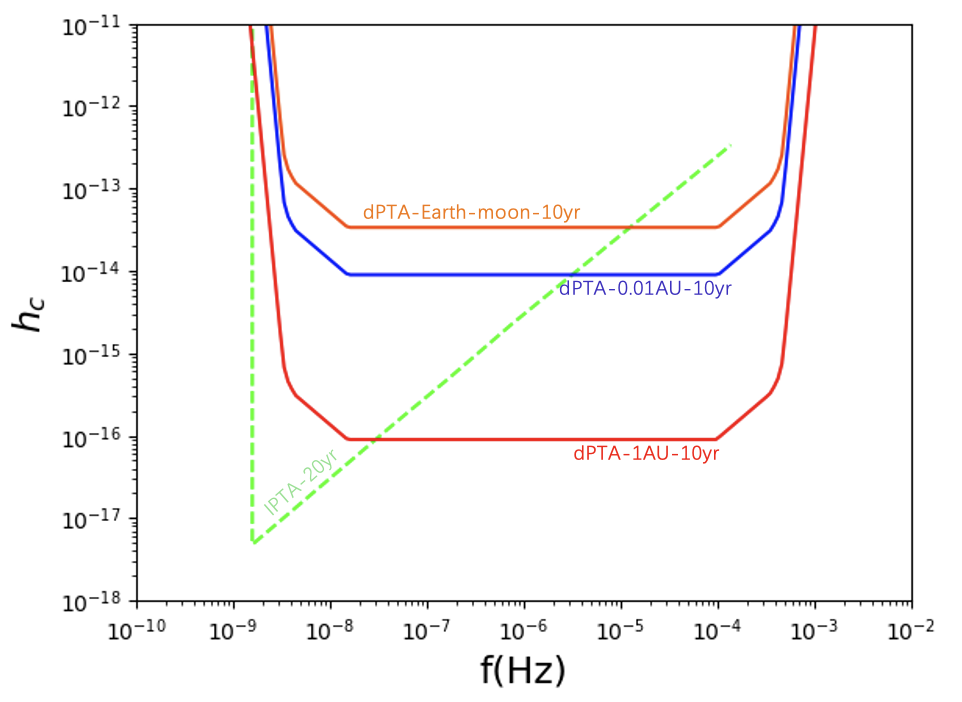}
\caption{Strain sensitivity curves for dPTA with observation time $T=10$ years. We select $D=3\times10^{-3}$AU (approximating the Earth-Moon distance, orange line), $D$=0.01AU (Earth-L2 distance, blue line) and $D$=1AU (Earth-L4 distance, red line). These are compared with the IPTA-20~\cite{Kuroda:2015owv,Antoniadis:2022pcn} sensitivity curve (green line). \justifying}
\label{h_c}
\end{figure}

\begin{figure}[h]
\captionsetup{justification=raggedright,singlelinecheck=false}
\includegraphics[scale=0.33]{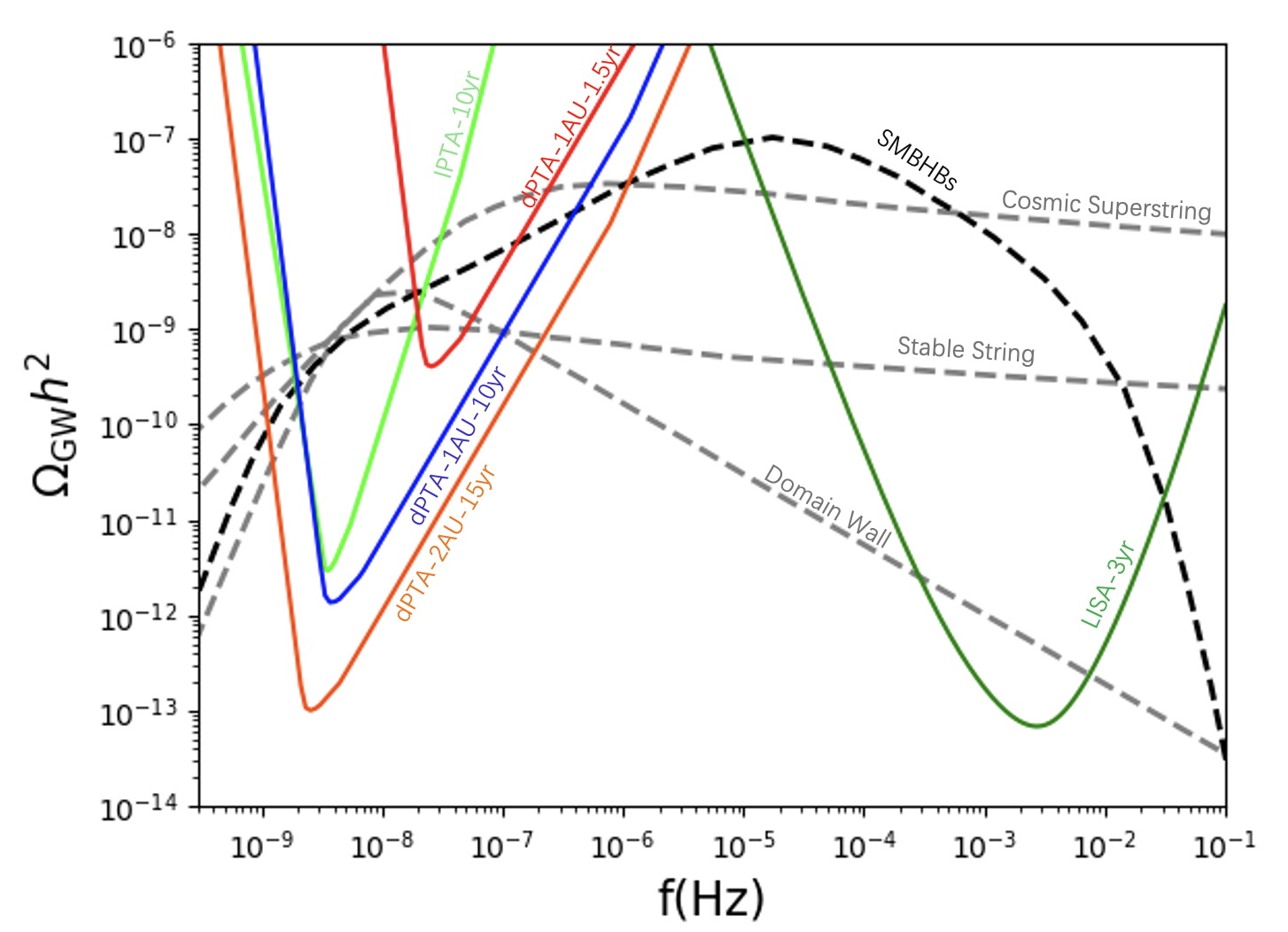}
\caption{\justifying
Power-law sensitivity curves for the dPTA, IPTA (lime line)~\cite{hobbs2010international} and LISA (green line)~\cite{Caprini:2019pxz}. For dPTA, we show configurations for 1AU baseline observed over 1.5 years (red line) and 10 years (blue line), and 2AU for 15 years (orange line).  We  compare them with some new physics spectral density curves (grey lines)~\cite{NANOGrav:2023hvm} and the SMBHBs spectral density curve (black line)~\cite{Bi:2023tib}.}
\label{sensitivecurve}
\end{figure}

Then,  we convert the energy density sensitivity curve to the strain sensitivity curve using the relation $\Omega_{\text{GW}}=\frac{2\pi^{2}f^{2}}{3H_{0}^{2}}h_{c}^{2}$ and present the results in Fig.~\ref{h_c}.
Our results demonstrate that $h_{c}\propto D^{-1}$. When the arm length increases, the $\mathcal{E}_{ab}(f,D,\hOmega)$ introduces a pronounced minimum on the sensitivity curve. Specifically for the nHz frequency band,  while optimizing the arm length can suppress the dPTA strain sensitivity curve by approximately eight orders of magnitude, the required baseline distance reaches astrophysical scales on the order of light-years.
Moreover, in Fig.~\ref{sensitivecurve}, we compare the dPTA energy density sensitivity curve with those of IPTA and LISA, with the new physical and supermassive black hole binaries (SMBHs) to examine its capability of detecting the GWB.

Thus, from Fig.~\ref{sensitivecurve}, we can find that the sensitivity of the dipole system with a 1AU  baseline not only reaches the level of IPTA in the nHz frequency band, but also extends into the $\mu$Hz range. Consequently, it helps to compensate for the lack of measurement for chirality and expand the frequency regime that is not covered by existing projects.

\section{Conclusion and Discussion}\label{section conclusion}

In this work, we construct a dPTA based on the PTA, designed to probe GWB by measuring the timing residual differences between two detectors. Through the analysis of the ORFs, we find that the dPTA exhibits sensitivity to the chirality of the GWB. Then, using the power-law integrated method to generate the sensitivity curves, we demonstrate that the dPTA has comparable sensitivity to both $I$ and $V$, with a detection band spanning from nHz to $\mu$Hz. 
Another similar approach for extending sensitivity into the $\mu$Hz range is FRB timing~\cite{Lu:2024yuo}. By measuring the differences in the arrival times of FRBs, a similar effect can be achieved. Since both methods rely on timing discrepancies, we can also establish correlations between FRB and pulsar signals, which may further enhance sensitivity.

Furthermore, numerous missions have recently been proposed across various GW detection projects.  In the mHz frequency band,  space-based GW detectors, such as LISA, Taiji and TianQin, can now probe the chirality and polarization through a joint network of detectors~\cite{Seto:2020zxw,Liu:2022umx,Chen:2024xzw,Chen:2024ikn,Chen:2024fto,Su:2025nkl}.  In the nHz frequency band,  although astrometry has been proposed as a complementary probe of the $V$ component~\cite{Book:2010pf,Liang:2023pbj,Jaraba:2025hay,Qin:2018yhy}, to which PTA are not sensitive in an isotropic background, these methods covering the $\mu$Hz frequency band remain lacking~\cite{Caliskan:2023cqm,Wang:2022sxn}. Therefore, the dPTA provides a promising approach for detecting both the intensity and chirality of the GWB across the nHz$\sim\mu$Hz frequency range. In future work, we plan to extend this framework to PTA-like systems, such as a dipole configuration for the recently proposed artificial precision timing array~\cite{Alves:2024ulc,Jiang:2025uga}, which is designed to detect mHz$\sim$decihertz GWB.

In conclusion, we consider that detecting chiral GWB can deepen our understanding of the early universe and astrophysical processes. Therefore, as a valuable complementary approach to a conventional PTA, the dPTA can not only extend the PTA detection band and broaden the scope of detectable sources, but also enhance its capability to test fundamental physics by detecting Stokes parameter $V$.

\begin{acknowledgments}
This work is supported by 
the National Natural Science Foundation of China (No.12588101, No.12375059), the National Key Research and Development Program of China (No.2023YFC2206200, No.2021YFA0718304).
\end{acknowledgments}

\appendix

\section{The cross-correlation for dPTA}\label{cc}
By using Eq.~\eqref{redshift}, the difference of redshift  between the two detectors for Pulsar-a can be expressed as
\begin{align}
&s_{a}(t) =  z_{1a}(t)-z_{2a}(t)       \\=& \int_{S^2} \frac{\dd\hat{\Omega} }{2} 
\Bigg[\frac{\hat{p}_{1a}^i \hat{p}_{1a}^j \Delta h_{ij}^{1a}(t,\hOmega)}{1 + \hat{\Omega} \cdot \hat{p}_{1a}} -\frac{\hat{p}_{2a}^i \hat{p}_{2a}^j\Delta h_{ij}^{2a}(t,\hOmega)}{1 + \hat{\Omega} \cdot \hat{p}_{2a}} \Bigg].  \nonumber
\end{align}
Here, $\Delta h_{ij}^{1a}(t,\hOmega)$ is the difference between the metric perturbations at the telescope-1 ($t_{1},x_{1}$) and at the Pulsar-a ($t_{a},x_{a}$), 
\begin{equation}\label{A2}
    \Delta h_{ij}^{1a}(t,\hOmega) = h_{ij}(t_{a},\hOmega)-h_{ij}(t_{1},\hOmega).
\end{equation}
Since the pulse emission time is the same, we take $t$ as the emission time here. When using Eq.~\eqref{GW fourier} and setting $t_{a}=t$, $t_{1}=t+L_{1a}$, given that 
$L_{1a}\gg D$, we can employ the following approximation
\begin{equation}\label{approx}
    \hp_{1a}=\frac{L_{1a}\hp_{1a}}{L_{1a}}=\frac{L_{a}}{L_{1a}}\hp_{a}+\frac{D}{2L_{1a}}\hat{D}\simeq \hp_{a}.
\end{equation}
Then, the Eq.~\eqref{A2} can be written as
\begin{align}
    \Delta h_{ij}^{1a} =& \sum_{\lambda} \int_{-\infty}^\infty \dd f \tilde{h}_{\lambda}(f, \hat{\Omega}) \, e^{\lambda}_{ij}(\hat{\Omega})e^{i2\pi f (t-L_{a}\hOmega\cdot\hat{p}_{a})} \nonumber\\
    &\times \left(1-e^{i 2\pi f L_{1a}(1 + \hat{\Omega} \cdot \hat{p}_{1a})} \right).
\end{align}
Thus, the signal $s_a(t)$ represents the difference in timing residuals of a pulsar $a$ measured by two telescopes.  The distance from the telescope to the millisecond pulsars is around several $\text{kpc}$, which is much greater than the distance between the two telescopes (around $1\text{AU}$). Consequently, we only need to consider the leading order term and obtain the redshift difference for Pulsar-a as
\begin{align}\label{approx1}
        s_{a}(t) =&\sum_{\lambda}\int_{S^2}  \dd\hat{\Omega} \int \dd f\frac{1}{2} \frac{\hp_{a}^{i} \hp^{j}_{a}e^{\lambda}_{ij}\tilde{h}_{\lambda}(f,\hOmega)}{1 + \hat{\Omega} \cdot \hat{p}_{a}} e^{i2\pi f (t-L_{a}\hOmega\cdot\hat{p}_{a})}
        \nonumber\\
        &\times \left[e^{i 2\pi f L_{2a}(1 + \hat{\Omega} \cdot \hat{p}_{2a})} -e^{i 2\pi f L_{1a}(1 + \hat{\Omega} \cdot \hat{p}_{1a})} \right]. 
    \end{align}
Under the approximation in Eq.~\eqref{approx},
the phase term in Eq.~\eqref{approx1} can be expressed as
\begin{align}
& e^{i2\pi f (t-L_{a}\hOmega\cdot\hat{p}_{a})} \left[e^{i 2\pi f L_{2a}(1 + \hat{\Omega} \cdot \hat{p}_{2a})} -e^{i 2\pi f L_{1a}(1 + \hat{\Omega} \cdot \hat{p}_{1a})}\right] 
\nonumber \\
\simeq & 2i e^{i 2 \pi f (t+L_{a})} \tsin\left[\pi f D(\hat{p}_{a}+\hOmega)\cdot\hat{D}\right].
\end{align}
The Fourier transform $\tilde{s}_{a}(f)$ of the redshift difference can then be extracted from $s_a(t)$ by applying the variable substitution $t=t+L_{a}$ in Eq.~\eqref{approx1}, which can be expressed as
\begin{equation}\label{signal_A}
    \tilde{s}_{a}(f) = \sum_{\lambda}\int \dd \hOmega  \mathcal{F}_{a}^{\lambda}(\hOmega) \tilde{h}_{\lambda}(f,\hOmega) \mathcal{E}_{a}(\hOmega). 
\end{equation}
The directional dependence part and phase part are
\begin{align}\label{F}
\begin{split}
    \mathcal{F}_{a}^{\lambda}(\hOmega) &=  \frac{\hp_{a}^{i} \hp^{j}_{a}e^{\lambda}_{ij}}{1 + \hat{\Omega} \cdot \hat{p}_{a}},
    \\
    \mathcal{E}_{a}(\hOmega) &= i \tsin[\pi f D(\hat{p}_{a}+\hOmega)\cdot \hat{D}].
\end{split}
\end{align}

The Fourier transform $\widetilde{s}(f)$ of the signal $s(t)$ is given by  Eq.~\eqref{signal_A}.  We then define the correlation function between the signal pairs. Following the functional form of  Eq.~\eqref{correlationS}, this correlation can be expressed as
\begin{equation}\label{ccccc}
\begin{split}
& \langle S_{ab}(\tau)\rangle = \int \dd t \langle s_{a}(t)s_{b}(t+\tau)\rangle \\
    &= \iint \dd f \dd f' \langle \tilde{s}_{a}(f) \tilde{s}^{*}_{b}(f') \rangle \delta_T(f-f')e^{-i 2\pi f' \tau}.
\end{split}
\end{equation}
Here, $\delta_{T}(f-f')$ is the finite-time approximation to the Dirac delta function $\delta(f-f')$, defined as
$\delta_{T}(f-f') = \frac{\tsin [\pi (f-f') T]}{\pi (f-f')}$, with the property that $\delta_{T}(0)=T$. Using Eq.~\eqref{signal_A}, the correlation in Eq.~\eqref{ccccc} can be written as
\begin{equation}
\begin{split}
 \langle \tilde{s}_{a}(f)& \tilde{s}_{b}^{*}(f') \rangle
=  \sum_{\lambda}\sum_{\lambda'}  \iint\dd\hat{\Omega} \dd\hat{\Omega'}
   \mathcal{E}_{a}({\hOmega})\mathcal{E}_{b}^{*}({\hOmega'}) \\    &\times 
 \langle\tilde{h}_{\lambda}(f,\hat\Omega)\tilde{h}_{\lambda'}^{*}(f',\hat\Omega')\rangle \mathcal{F}_{a}^{\lambda}(\hOmega) \mathcal{F}_{b}^{\lambda'}(\hOmega') .
\end{split}
\end{equation}
Upon expanding the above expression, we get
\begin{equation}\label{signal_expaned_A}
\begin{split}
 \langle \tilde{s}_{a}(f)& \tilde{s}_{b}^{*}(f') \rangle
=  \iint\dd\hat{\Omega} \dd\hat{\Omega'}
   \mathcal{E}_{a}({\hOmega})\mathcal{E}_{b}^{*}({\hOmega'})\times \\    
   &\big[
  \mathcal{F}_{a}^{+}(\hOmega) \mathcal{F}_{b}^{+}(\hOmega')\langle\tilde{h}_{+}(f,\hat\Omega)\tilde{h}_{+}^{*}(f',\hat\Omega')\rangle \\
  &+
 \mathcal{F}_{a}^{+}(\hOmega) \mathcal{F}_{b}^{\times}(\hOmega')\langle\tilde{h}_{+}(f,\hat\Omega)\tilde{h}_{\times}^{*}(f',\hat\Omega')\rangle \\
  &+
 \mathcal{F}_{a}^{\times}(\hOmega) \mathcal{F}_{b}^{+}(\hOmega')\langle\tilde{h}_{\times}(f,\hat\Omega)\tilde{h}_{+}^{*}(f',\hat\Omega')\rangle \\
  &+
  \mathcal{F}_{a}^{\times}(\hOmega) \mathcal{F}_{b}^{\times}(\hOmega')\langle\tilde{h}_{\times}(f,\hat\Omega)\tilde{h}_{\times}^{*}(f',\hat\Omega')\rangle \big].
\end{split}
\end{equation}
Here, the ensemble average is given by
\begin{align}
 &\langle \tilde{h}_{\lambda}(f,\hat\Omega)\tilde{h}_{\lambda'}^{*}(f',\hat\Omega')\rangle \nonumber\\
 =  &\left[\begin{array}{cc}
     \langle\tilde{h}_{+}(f,\hat\Omega)\tilde{h}_{+}^{*}(f',\hat\Omega')\rangle & \langle\tilde{h}_{+}(f,\hat\Omega)\tilde{h}_{\times}^{*}(f',\hat\Omega')\rangle \\
     \langle\tilde{h}_{\times}(f,\hat\Omega)\tilde{h}_{+}^{*}(f',\hat\Omega')\rangle  & \langle\tilde{h}_{\times}(f,\hat\Omega)\tilde{h}_{\times}^{*}(f',\hat\Omega')\rangle 
     \end{array}\right]\nonumber\\
 =   &\frac{1}{8\pi}\delta(f-f')\delta(\hOmega-\hOmega') 
     \left[\begin{array}{cc}
     I(f) & -iV(f) \\
     iV(f) & I(f)
     \end{array}\right].
\end{align}
Substituting the components from the above expression into  Eq.~\eqref{signal_expaned_A}, gives
\begin{equation}\label{signal_expaned_A}
\begin{split}
 \langle \tilde{s}_{a}(f)& \tilde{s}_{b}^{*}(f') \rangle
=  \frac{\delta(f-f')}{8\pi}\int\dd\hat{\Omega}
   \mathcal{E}_{a}({\hOmega})\mathcal{E}_{b}^{*}({\hOmega'})\times \\    
   &\big[
  \mathcal{F}_{a}^{+} \mathcal{F}_{b}^{+}I(f) 
  -
  \mathcal{F}_{a}^{+}\mathcal{F}_{b}^{\times}iV(f) \\
  &+
  \mathcal{F}_{a}^{\times} \mathcal{F}_{b}^{+}iV(f) 
  +
  \mathcal{F}_{a}^{\times} \mathcal{F}_{b}^{\times}I(f) \big].
\end{split}
\end{equation}
Then, extracting components with identical Stokes parameters,
\begin{equation}
    \begin{split}
& \langle \tilde{s}_{a}(f) \tilde{s}_{b}^{*}(f') \rangle  = \frac{\delta(f-f')}{2}\int\dd\hat{\Omega} \mathcal{E}_{a}({\hOmega})\mathcal{E}_{b}^{*}({\hOmega})\times
        \\
        &\left[\frac{I(f)}{4\pi}\left(\mathcal{F}_{a}^{+}\mathcal{F}_{b}^{+}+\mathcal{F}_{a}^{\times}\mathcal{F}_{b}^{\times}\right) - \frac{iV(f)}{4\pi}(\mathcal{F}_{a}^{+}\mathcal{F}_{b}^{\times}-\mathcal{F}_{a}^{\times}\mathcal{F}_{b}^{+})\right].
    \end{split}
\end{equation}
The above expression can be rewritten as
\begin{equation}\label{A15}
    \begin{split}
&\langle \tilde{s}_{a}(f) \tilde{s}_{b}^{*}(f') \rangle  = \frac{1}{2}\delta(f-f')\times \\
&\big\{
I(f)\big[\int 
       \frac{\dd\hat{\Omega}}{4\pi}\mathcal{E}_{a}({\hOmega})\mathcal{E}_{b}^{*}({\hOmega})\left(\mathcal{F}_{a}^{+}\mathcal{F}_{b}^{+}+\mathcal{F}_{a}^{\times}\mathcal{F}_{b}^{\times}\right)\big] \\ &+V(f)\big[-i\int\frac{\dd\hat{\Omega}}{4\pi}\mathcal{E}_{a}({\hOmega})\mathcal{E}_{b}^{*}({\hOmega})(\mathcal{F}_{a}^{+}\mathcal{F}_{b}^{\times}-\mathcal{F}_{a}^{\times}\mathcal{F}_{b}^{+})\big]\big\}.
    \end{split}
\end{equation}
It follows that the correlation comprises the Stokes parameters and a spatial integrals. We then define the ORFs of the dPTA system as follows
\begin{align}\label{A16}
    \begin{split}
       \widetilde{\Gamma}_{ab}^{I}(f) &= \int \frac{\dd \hOmega}{4\pi} \left(\mathcal{F}_{a}^{+}\mathcal{F}_{b}^{+}+\mathcal{F}_{a}^{\times}\mathcal{F}_{b}^{\times}\right)\mathcal{E}_{a}({\hOmega})\mathcal{E}_{b}^{*}({\hOmega}),\\
       \widetilde{\Gamma}_{ab}^{V}(f) &=  -i\int \frac{\dd \hOmega}{4\pi} (\mathcal{F}_{a}^{+}\mathcal{F}_{b}^{\times}-\mathcal{F}_{a}^{\times}\mathcal{F}_{b}^{+}) \mathcal{E}_{a}({\hOmega})\mathcal{E}_{b}^{*}({\hOmega}).
    \end{split}
\end{align}
Note that in \eqref{A15}, we have moved the imaginary factor $i$ of $iV(f)$
into the integral, which is the origin of the imaginary nature of $\widetilde{\Gamma}^{V}_{ab}$.
Substituting Eq.~\eqref{F} into Eq.~\eqref{A16}, we can obtain Eq.~\eqref{GammaI} and Eq.~\eqref{GammaV}. From Eq.~\eqref{F}, we can observe that spatial part $(\mathcal{F}_{a}^{+}\mathcal{F}_{b}^{\times}-\mathcal{F}_{a}^{\times}\mathcal{F}_{b}^{+})$ in $\widetilde{\Gamma}^{V}_{ab}$ is real. The phase part, that $\mathcal{E}_{a}({\hOmega})\mathcal{E}_{b}^{*}({\hOmega})=\tsin[\pi f D(\hat{p}_{a}+\hOmega)\cdot \hat{D}]\tsin[\pi f D(\hat{p}_{b}+\hOmega)\cdot \hat{D}]$ is also real. Therefore, the imaginary factor $i$ in $\widetilde{\Gamma}_{ab}^{V}(f)$ stems from the absorption of the imaginary factor $i$ from $\langle \tilde{h}_{\lambda}h_{\lambda'}^{*}\rangle$ into the spatial integral. In summary, due to the definition of Eq.~\eqref{A15}, the imaginary factor of $iV(f)$ is absorbed into $\widetilde{\Gamma}_{ab}^{V}$, thereby inducing the imaginary nature of $\widetilde{\Gamma}_{ab}^{V}$. 
Finally,  inserting $\langle \tilde{s}_{a}(f) \tilde{s}_{b}^{*}(f') \rangle
= \frac{1}{2}\delta(f-f')\left[I(f)\widetilde{\Gamma}_{ab}^{I}(f)+V(f)\widetilde{\Gamma}_{ab}^{V}(f)\right]$ into Eq.~\eqref{ccccc}, we can obtain an equation similar to Eq.~\eqref{cor}, which can be expressed as 
\begin{equation}
\begin{split}
\langle S_{ab}(\tau)\rangle= \frac{T}{2} \int\dd f&\big[I(f)\widetilde{\Gamma}_{ab}^{I}(f)\tcos(2\pi f \tau)
\\
&-V(f) i \widetilde{\Gamma}_{ab}^{V}(f)\tsin(2\pi f \tau)\big].
\end{split}
\end{equation}
Similar to PTA analyses, the above expression is real. However, in the dPTA case, we have $i \widetilde{\Gamma}_{ab}^{V}\neq 0$, , which allows us to obtain the response to $V(f)$.

\section{Signal to noise radio}\label{SNRSNR}

In this section, we present the derivation of the SNR and the optimal filter function required to maximize it. We follow the derivation process in~\cite{Allen:1997ad}. In practical measurements, noise is inevitably present. The total signal can be expressed as $s_{\text{total}}(t)=s(t)+n(t)$, where $n(t)$ depends on the noise model. Assuming that the noises are independent, the cross-correlation can be defined as~\cite{Kato:2015bye}
\begin{equation}
    \begin{split}
       S_{ab}  &= \iint \dd t  \dd t'  s_{a}(t)K_{ab}(t-t')s_{b}(t') 
       \\ 
        &= \iint \dd f  \dd f' \delta_{T}(f-f')\tilde{s}_{a}(f)\widetilde{K}_{ab}(f')\tilde{s}_{b}^{*}(f').
    \end{split}
\end{equation}
Assuming that the noise is stationary, Gaussian, and uncorrelated between detectors, the average of the cross-correlation can be rewritten as
\begin{equation}\label{AS}
    \langle S_{ab} \rangle = \iint \dd f  \dd f' \delta_{T}(f-f')\widetilde{K}_{ab}(f)
     \langle \tilde{s}_{a}(f) \tilde{s}_{b}^{*}(f') \rangle.
\end{equation}
Here, $\widetilde{K}_{ab}(f)$ denotes the filter function.
For the cross-correlation between signals from Pulsar-a and Pulsar-b, the optimal filter function that maximizes the SNR can be written as
\begin{equation}
    \widetilde{K}_{ab}(f) = \frac{I(f)\widetilde{\Gamma}_{ab}^{I}(f)+V(f)\widetilde{\Gamma}_{ab}^{ V}(f)}{P(f)P(f)}.
\end{equation}

SNR is defined as $\text{SNR}=\mu/\sigma$, where $\mu$ and $\sigma^{2}$ are the mean value and variance of the cross-correlation signal $S_{ab}$. The mean value $\mu=\langle S_{ab} \rangle$ is given by ~\eqref{AS}.
When the pulsars share the same white noise, the variance of the cross-correlation signal can be written as
\begin{equation}
    \sigma^{2} = \frac{T}{4} \int \dd f P^{2}(f) |\widetilde{K}_{ab}(f)|^{2}.
\end{equation}
According to Eq. (3.69) in \cite{Allen:1997ad},  the mean value and variance can be expressed as two positive definite inner products:
\begin{align}
    \begin{split}
& \mu =  T \left(\widetilde{K}_{ab}(f),\frac{\hat{\Gamma}_{ab}(f)}{P(f) P(f)}\right) ,\\
&\sigma^{2} = \frac{T}{4}\left(\widetilde{K}_{ab}(f), \widetilde{K}_{ab}(f)\right).
    \end{split}
\end{align}
Thus, the maximum SNR can be written as:
\begin{equation}\label{SNRinner}
    \text{SNR}^{2} = \frac{\mu^{2}}{\sigma^{2}} = 4T\frac{ \left(\widetilde{K}_{ab}(f),\frac{\hat{\Gamma}_{ab}(f)}{P(f) P(f)}\right)^{2}}{\left(\widetilde{K}_{ab}(f), \widetilde{K}_{ab}(f)\right)}.
\end{equation}
The Cauchy–Schwarz inequality states that for any vectors (or functions) $A$ and $B$ in an inner product space, $| (A, B) |^2 \leq (A, A) \, (B, B)$, with equality if and only if $A$ and $B$ are linearly dependent. That is, there exists a complex number ${\mC}$ such that $A= {\mC} B$. Therefore, the filter function that maximizes the SNR can be expressed as:
\begin{equation}
    \widetilde{K}_{ab}(f) =  \frac{{\mC}\,\hat{\Gamma}_{ab}(f)}{P(f) P(f)},
\end{equation}
and the maximum SNR in Eq. \eqref{SNRinner} is given by
\begin{equation}
    \begin{split}
        \text{SNR}^{2} &= 4T  \left(\frac{\hat{\Gamma}_{ab}(f)}{P(f) P(f)},\frac{\hat{\Gamma}_{ab}(f)}{P(f) P(f)}\right)
        \\
        &=T \int \dd f \frac{\left[I(f)\widetilde{\Gamma}_{ab}^{I}(f)+V(f)\widetilde{\Gamma}_{ab}^{ V}(f)\right]^{2}}{P(f) P(f)}.
    \end{split}
\end{equation}

When extending to the case of multiple pulsars, each pair can be cross-correlated. Therefore, the optimal filter function $\widetilde{K}_{ab}(f)$ takes the form of a matrix, with each element corresponding to the optimal filter function for a specific cross-correlation pair. The matrix elements can be written as: $\widetilde{K}_{ab}(f) = {\mC}\, \hat{\Gamma}_{ab}(f)/P^{2}(f)$.
Just as in Eq.~\eqref{SNRinner}, the SNR can also be written in the form of an inner product of the matrix
\begin{equation}
    \begin{split}
        \text{SNR}^{2} &= 4T\int \dd f \text{Tr} \left[\widetilde{K}_{ab}(f)\widetilde{K}_{ab}^{\dagger}(f)\right]\frac{P(f)^{2}}{{\mC}^{2}} 
        \\
        &=T \int \dd f \sum_{a,b} \frac{\left[I(f)\widetilde{\Gamma}_{ab}^{I}(f)+V(f)\widetilde{\Gamma}_{ab}^{ V}(f)\right]^{2}}{P(f) P(f)}.
    \end{split}
\end{equation}
Converted to the single-sided spectrum in the observable frequency band, the maximum SNR can be written as:
\begin{equation}
    \begin{split}
        \text{SNR}^2=2T \int^{f_{\text{max}}}_{f_{\text{min}}} \dd f \sum_{a,b} \frac{\left[I(f)\widetilde{\Gamma}_{ab}^{I}(f)+V(f)\widetilde{\Gamma}_{ab}^{ V}(f)\right]^{2}}{P(f) P(f)}.
    \end{split}
\end{equation}
Thus, the SNR depends on the coupling between $I$ and $V$. 

To clearly identify the parity-violating component, one must to separate the $I$ and $V$ components from the cross-correlation of the detected signals. The SNR can be written in quadratic form in terms of $I$ and $V$:
\begin{align}
\begin{split}
    \text{SNR}^{2} &= \int \dd f \,
    {\Theta}^\mathrm{T}(f)\mathcal{F}(f){\Theta}(f), 
    \\
    {\Theta}^\mathrm{T} &=\left[\begin{matrix}I(f) & V(f)\end{matrix}\right].
\end{split}
\end{align}
Here, $\mathcal{F}(f)$ is the Fisher information density matrix, whose explicit form is:
\begin{equation}
\mathcal{F}(f) = \frac{2T}{P(f) P(f)} \sum_{a,b}
\left[\begin{matrix}
(\widetilde{\Gamma}_{ab}^{I})^{2} & \widetilde{\Gamma}_{ab}^{I} \widetilde{\Gamma}_{ab}^{V} \\
\widetilde{\Gamma}_{ab}^{I} \widetilde{\Gamma}_{ab}^{V} & (\widetilde{\Gamma}_{ab}^{V})^2 
\end{matrix}\right].
\end{equation}
For convenience in the derivation, we write the Fisher matrix in a simplified form:
\begin{equation}
    \mathcal{F}(f) = \left[\begin{matrix}
A(f) & B(f) \\
B(f) & C(f) 
\end{matrix}\right].
\end{equation}
The effective SNR is related to the diagonal elements of the covariance matrix, therefore, we need to compute the inverse of the Fisher matrix:
\begin{equation}
\mathcal{F}^{-1}(f) = \frac{1}{A(f) C(f) - B(f)^2}
\left[\begin{matrix}
C(f) & -B(f) \\
-B(f) & A(f)
\end{matrix}\right].
\end{equation}
Then the effective SNRs can be written as
\begin{align}
    \text{SNR}_{I}^{2} = \int \dd f \frac{A(f)C(f)-B(f)^{2}}{C(f)}I^{2},
    \\
    \text{SNR}_{V}^{2} = \int \dd f \frac{A(f)C(f)-B(f)^{2}}{A(f)}V^{2}.
\end{align}
Substituting the explicit forms of the matrix elements back yields formulas:

\begin{align}
 \text{SNR}_{I}^{2} &=  
    2T\int^{f_{\text{max}}}_{f_{\text{min}}} \dd f \left[  \widetilde{\Gamma}^{ I}_{\text{eff}}(f)\right]^2\frac{I(f)^{2}}{P(f)^2},\label{SNR_I}
    \\
    \text{SNR}_{V}^{2} &=  
    2T\int^{f_{\text{max}}}_{f_{\text{min}}} \dd f \left[  \widetilde{\Gamma}^{ V}_{\text{eff}}(f)\right]^2\frac{V(f)^{2}}{P(f)^2}.\label{SNR_V}
\end{align}

We have defined two effective ORFs accordingly, which can be expressed as
\begin{align}
    \widetilde{\Gamma}^{I}_{\text{eff}}(f)&=\sqrt{\sum_{a,b}\widetilde{\Gamma}_{ab}^{ I}(f)^2-\frac{\left[\sum_{a,b}\widetilde{\Gamma}_{ab}^{I}(f) \widetilde{\Gamma}_{ab}^{ V}(f)\right]^{2}}{ \sum_{a,b}\widetilde{\Gamma}_{ab}^{V}(f)^2}},
    \\
    \widetilde{\Gamma}^{V}_{\text{eff}}(f)&=\sqrt{\sum_{a,b}\widetilde{\Gamma}_{ab}^{ V}(f)^2-\frac{\left[\sum_{a,b}\widetilde{\Gamma}_{ab}^{I}(f) \widetilde{\Gamma}_{ab}^{ V}(f)\right]^{2}}{ \sum_{a,b}\widetilde{\Gamma}_{ab}^{I}(f)^2}}.
\end{align}
Since our PTA contains many pulsars that effectively average over sky positions, we simplify the calculation by averaging the geometric parts of the ORFs.  Accordingly, in the ideal case of $N$ pulsars, the effective ORFs for $I$ and $V$ can be rewritten as:
\begin{align}
    \widetilde{\Gamma}^{I}_{\text{eff}}(f) = \sqrt{\frac{N(N-1)}{2}\left(\overline{\widetilde{\Gamma}_{ab}^{ I}(f)^2}-\frac{\left[\overline{\widetilde{\Gamma}_{ab}^{I}(f) \widetilde{\Gamma}_{ab}^{ V}(f)}\right]^{2}}{ \overline{\widetilde{\Gamma}_{ab}^{V}(f)^2}}\right)},
    \\
     \widetilde{\Gamma}^{V}_{\text{eff}}(f) =\sqrt{ \frac{N(N-1)}{2}\left(\overline{\widetilde{\Gamma}_{ab}^{ V}(f)^2}-\frac{\left[\overline{\widetilde{\Gamma}_{ab}^{I}(f) \widetilde{\Gamma}_{ab}^{ V}(f)}\right]^{2}}{ \overline{\widetilde{\Gamma}_{ab}^{I}(f)^2}}\right)}.
\end{align} 
Finally, by employing the power-law integrated method, we can obtain the sensitivity curves for the Stokes parameters $I$ and $V$.
The sensitivity curves for the Stokes parameters can be written as
\begin{equation}
   \Omega_{\text{GW}}^{I,V}(f) = \underset{\beta}{\text{max}}\left[f^{\beta}\cdot\frac{\text{SNR}_{I,V}}{\sqrt{2T}}\Bigg[\int^{f_{\text{max}}}_{f_{\text{min}}}\dd f\Big(\frac{f^{\beta}}{\Omega^{ I, V}_{\text{eff}}}\Big)^{2}\Bigg]^{-\frac{1}{2}}\right].
\end{equation}
Here, $\beta$ denotes the power-law index for the different power-law Gaussian stochastic GWB and the effective energy density $\Omega^{I,V}_{\text{eff}}$ can be expressed as 
\begin{equation}
    \Omega^{I,V}_{\text{eff}} = \frac{4\pi^{2}}{3H_{0}^{2}}f^{3}\left[\frac{\widetilde{\Gamma}^{I, V}_{\text{eff}}(f)^{2}}{P(f)P(f)} \right]^{-\frac{1}{2}}.
\end{equation}
When the GW wavelength is much larger than the baseline $D$, we have $\tsin [\pi f D (\hat{p}_{a}+\hOmega)\cdot\hat{D} ]\simeq\pi f D (\hat{p}_{a}+\hOmega)\cdot\hat{D}$, implying that $\Omega^{I,V}_{\text{GW}}\propto D^{-2}$ and $h_c\propto D^{-1}$. However, as $D$ increases, the approximation condition is no longer satisfied. The periodicity of the function $\tsin [\pi f D (\hat{p}_{a}+\hOmega)\cdot\hat{D} ]$ will  lead to the emergence of a maximum value for $\widetilde{\Gamma}^{I,V}_{\text{eff}}$.

\bibliography{refs.bib}

\end{document}